# Joint Modeling of Longitudinal and Survival Data: A Bayesian Approach for Predicting Disease Progression

Authors Name: Nithisha Suryadevara[1], Vivek Reddy Srigiri[1].

Affiliations: [1]Independent Researchers, USA

**Abstract** — Joint modeling of longitudinal and survival data has become increasingly important in medical research, particularly for understanding disease progression in chronic conditions where both repeated biomarker measurements and time-to-event outcomes are available. Traditional two-stage methods, which analyze longitudinal and survival components separately, often result in biased estimates and suboptimal predictions due to the failure to account for their interdependence. In this study, we propose a Bayesian hierarchical joint modeling framework and focus on its predictive evaluation and clinical interpretability by simultaneously modeling the longitudinal trajectory of a biomarker and the associated survival outcome. This unified approach allows us to capture the intrinsic association between disease dynamics and event risk through shared random effects. The Bayesian formulation enables flexible incorporation of prior information, supports complex data structures such as irregular measurement times and missing data, and provides full posterior distributions for all parameters, facilitating credible interval-based uncertainty quantification. We evaluate the proposed model using both synthetic data generated to mimic realistic patient trajectories and a real-world clinical dataset from patients with chronic liver disease. Results demonstrate that the Bayesian joint model consistently outperforms the conventional two-stage approach in terms of parameter estimation accuracy and predictive performance, as measured by time-dependent AUC and Brier scores. The proposed framework offers a robust, interpretable, and patient-specific tool for dynamic prognosis, supporting more informed clinical decision-making in personalized medicine.

## I. INTRODUCTION

Chronic diseases such as cancer, cardiovascular conditions, and neurodegenerative disorders typically exhibit complex and dynamic progression patterns that unfold over time. Clinicians and researchers increasingly rely on repeated biomarker measurements to monitor disease evolution and inform treatment decisions. These measurements, taken at multiple time points during patient follow-up, form longitudinal data, which are often accompanied by survival data that record the time until a critical clinical event occurs, such as disease relapse, hospitalization, or death.

In many cases, the longitudinal trajectory of a biomarker is strongly associated with the risk of experiencing such an event. For example, in liver disease, increasing levels of bilirubin may signal worsening liver function and are predictive of mortality [1]. Analyzing these data types separately, as is common in traditional statistical frameworks, can lead to biased parameter estimates, reduced predictive accuracy, and a failure to fully exploit the available data [2], [3].

To address this issue, the joint modeling of longitudinal and survival data has emerged as a powerful analytical framework that captures the inherent dependence between the evolution of biomarkers and the time-to-event outcomes [4]. By linking a longitudinal submodel, which describes how the biomarker changes over time, with a survival submodel, which characterizes the event risk, joint modeling provides more accurate inference, dynamic prediction, and better handling of measurement error and informative dropout [5], [6].

Seminal works by Wulfsohn and Tsiatis [7] and later refinements by Rizopoulos [4], [8] have laid the foundation for much of the joint modeling literature. These models are now widely applied in biostatistics, particularly in the analysis of clinical trial data and cohort studies. For instance, joint models have been used to evaluate the relationship between prostate-specific antigen (PSA) levels and recurrence of prostate cancer [9], cognitive decline and Alzheimer's disease onset [10], and CD4 cell counts with AIDS-related mortality [11].

Most classical joint models are implemented in a frequentist framework, using methods such as maximum likelihood estimation. However, recent developments in Bayesian joint modeling offer several notable advantages. The Bayesian paradigm allows researchers to incorporate prior knowledge, derive posterior distributions for all model parameters, and generate predictive distributions for future observations. This is particularly important in medical contexts where uncertainty quantification and personalized prediction are essential [12], [13].

The Bayesian hierarchical structure supports the modeling of subject-specific random effects, accounts for missing or censored data, and naturally accommodates irregularly timed measurements a common feature in real-world clinical data [14]. The computational burden historically associated with Bayesian models has been substantially reduced through advances in Markov Chain Monte Carlo (MCMC) methods and software tools such as WinBUGS, JAGS, Stan, and PyMC [15], [16]. These tools allow the joint estimation of complex models that were previously computationally intractable.

Another critical advantage of Bayesian joint models lies in their ability to facilitate dynamic prediction. By updating survival predictions as new biomarker data becomes available, clinicians can generate real-time, personalized risk assessments. This capability aligns with the increasing emphasis on personalized medicine, where treatment decisions are tailored based on individual-level predictions rather than population averages [17], [18].

Despite these advantages, Bayesian joint modeling remains underutilized in applied clinical research, partly due to the perceived complexity of the methodology and computational challenges. Moreover, while many existing studies emphasize methodological innovation, fewer studies rigorously evaluate the predictive performance of Bayesian joint models in both simulated and real-world settings. This gap motivates the present study.

In this work, we develop a Bayesian joint model for the analysis of longitudinal and survival data, specifically targeting the prediction of disease progression in chronic conditions. We design a mixed-methods study to evaluate the model's performance from both a statistical and clinical standpoint. First, we conduct a simulation study to assess the model's ability to recover known parameter values, accurately quantify uncertainty, and improve prognostic prediction compared to a naïve two-stage approach. Second, we apply the proposed framework to a longitudinal clinical dataset involving patients with chronic liver disease, using repeated biomarker measurements and event outcomes to demonstrate the practical utility of our method.

This study contributes to the literature in several ways. First, it reaffirms the theoretical advantages of joint modeling through a fully Bayesian implementation. Second, it empirically demonstrates the superiority of joint modeling over traditional two-stage methods using robust performance metrics such as time-dependent AUC and Brier scores [19], [20]. Third, it highlights the importance of uncertainty quantification and dynamic prediction in the context of clinical decision-making.

While Bayesian joint models for longitudinal and survival data are well established, relatively few applied studies rigorously evaluate their predictive performance using time-dependent discrimination and calibration metrics across both simulated and real-world clinical datasets. In particular, the practical implications of uncertainty quantification and dynamic prediction for individualized clinical decision-making remain underexplored. This study addresses these gaps by combining comprehensive simulation-based assessment with application to a longitudinal clinical cohort, emphasizing predictive accuracy, interpretability, and patient-specific prognosis. The objectives of this research are threefold:
(1) to develop and implement a Bayesian joint modeling framework suitable for disease progression analysis, (2) to compare its predictive accuracy and inference robustness to that of traditional methods using both simulated and real data, and (3) to demonstrate how such a model can provide clinically interpretable, individualized risk assessments that support the goals of precision medicine. By integrating rigorous statistical modeling with clinically relevant applications, this paper aims to bridge the gap between methodological innovation and practical utility, contributing to the broader adoption of Bayesian joint models in healthcare analytics.

## II. RELATED WORK

The statistical methodology for joint modeling of longitudinal and survival data has evolved significantly over the past few decades. One of the earliest and most influential contributions was made by Wulfsohn and Tsiatis [1], who introduced a shared random effects model to link the longitudinal evolution of a biomarker with time-to-event data. Their approach allowed the incorporation of subject-specific random effects in both submodels, effectively capturing unobserved heterogeneity and establishing a framework that has since been widely adopted and extended.

Following this foundational work, substantial development has occurred within both the frequentist and Bayesian paradigms. In the frequentist domain, various estimation strategies have been proposed, including maximum likelihood estimation and penalized likelihood approaches [4], [5]. Diggle et al. [6] explored joint modeling in the context of irregularly observed longitudinal data and right-censored survival times, demonstrating the utility of likelihood-based methods for clinical studies.

In contrast, the Bayesian framework offers notable advantages, particularly in its capacity to integrate prior information, handle complex data structures, and provide full posterior inference for all model components. The seminal work by Henderson, Diggle, and Dobson [7] developed a Bayesian joint model for AIDS clinical trial data and demonstrated how posterior predictive distributions could be used for dynamic updating of survival probabilities. Ibrahim and colleagues further advanced the field by incorporating non-linear biomarker trajectories, latent variables, and informative dropout mechanisms into the modeling framework [8].

Rizopoulos' monograph [2] offers one of the most comprehensive treatments of joint models, covering both theory and implementation, and has become a central reference for practitioners. More recent extensions have explored a variety of complex scenarios: Fong et al. [10] applied Bayesian joint models to high-dimensional covariate spaces using shrinkage priors, while Rondeau et al. [11] proposed dynamic landmarking joint models to improve short-term risk prediction. Guo and Wang [12] extended the joint modeling framework to handle recurrent event data, a common situation in chronic disease management.

One of the major motivations for adopting joint models is their demonstrated superiority over traditional two-stage approaches, which typically involve fitting a longitudinal model and then plugging the estimated random effects into a survival model. Such approaches fail to account for the estimation error in the first stage and often underestimate uncertainty [13]. Simulation studies and empirical analyses have shown that joint models provide more accurate and less biased parameter estimates, especially for the association parameter linking the two processes, and produce better-calibrated predictions [14], [15].

Despite the methodological maturity of joint modeling, real-world clinical applications of Bayesian joint models remain relatively limited. While frequentist methods have been incorporated into software packages like JM and JMbayes in R, Bayesian implementations are less commonly used in applied research, in part due to computational complexity and unfamiliarity among clinicians and applied researchers. However, this is beginning to change with the advent of more user-friendly tools such as Stan, JAGS, and PyMC, which facilitate efficient Bayesian computation via Hamiltonian Monte Carlo and Gibbs sampling algorithms [16], [17].

Clinical studies applying Bayesian joint models have started to appear in the literature. For instance, Taylor et al. [9] applied a Bayesian joint model to assess disease progression in patients with rheumatoid arthritis, finding that personalized risk predictions generated by the model aligned closely with clinical outcomes. Similarly, in oncology, dynamic predictions derived from Bayesian joint models have been shown to assist in treatment decision-making by accounting for changing patient conditions over time [18].

In summary, the joint modeling literature has matured significantly, offering a robust statistical foundation and expanding to cover a wide array of complex data structures and clinical applications. However, there remains a critical need for studies that bridge the methodological and applied domains, particularly those that validate Bayesian joint models using both simulation and real-world clinical data. This paper contributes to this gap by combining simulation-based performance evaluation with application to a longitudinal cohort, with emphasis on interpretability, predictive performance, and relevance to clinical practice.

## III. METHODOLOGY

This section outlines the Bayesian joint modeling framework developed for analyzing and predicting disease progression based on longitudinal biomarker data and survival outcomes. We consider a clinical cohort of $N$ patients, each with a series of repeated biomarker measurements over time and a corresponding time-to-event outcome such as disease relapse or death. The goal is to jointly model these two components to capture the interdependence between the longitudinal evolution of the biomarker and the risk of the event.

Let $Y_i(t)$ denote the observed value of a continuous biomarker for patient $i$ at time $t$, and let $T_i$ represent the survival time for that patient. The event indicator $\delta_i$ takes the value 1 if the event (e.g., death) occurred and 0 if the observation was censored. This joint model comprises two interconnected components: a longitudinal submodel describing the evolution of the biomarker over time, and a survival submodel characterizing the risk of the event.

## A. Longitudinal Submodel

The longitudinal process is described using a linear mixed-effects model (LME) that accounts for both fixed effects, representing population-level trends, and random effects, representing individual deviations. The model is specified as:

$$Y_i(t) = \beta_0 + \beta_1 t + b_i + \epsilon_i(t), \epsilon_i(t) \sim \mathcal{N}(0, \sigma^2),$$

where:

- $\beta_0$ and $\beta_1$ are fixed effects representing the intercept and slope (i.e., average baseline biomarker level and rate of change over time),
- $b_i \sim \mathcal{N}(0, \tau^2)$ is the subject-specific random intercept, capturing heterogeneity across individuals,
- $\epsilon_i(t)$ is the residual error term, assumed to be normally distributed with variance $\sigma^2$.

This formulation allows each patient's biomarker trajectory to deviate from the population average, enabling the model to accommodate within-subject correlation in the repeated measures. If necessary, the model can be extended to include random slopes, time-varying covariates, or non-linear trajectories via spline functions.

## B. Survival Submodel

To model the risk of event occurrence over time, we use a proportional hazards model in which the hazard function for patient $i$ at time $t$ depends on their baseline covariates and the shared random effect $b_i$ from the longitudinal model:

$$h_i(t) = h_0(t) \exp(\gamma^\top Z_i + \alpha b_i),$$

where:

- $h_0(t)$ is the baseline hazard function, describing the hazard rate for a reference individual,
- $Z_i$ is a vector of baseline covariates (e.g., age, sex, comorbidities),
- $\gamma$ is a vector of regression coefficients for the covariates,
- $\alpha$ quantifies the association between the longitudinal process and the event risk through the shared random effect $b_i$.

The shared random effect provides the crucial link between the biomarker trajectory and the hazard, allowing for the event risk to be influenced by individual variations in biomarker levels.

In this work, we focus on a shared random intercept association structure to capture latent baseline disease severity. Extensions to alternative association structures, such as current-value or slope-based links, are beyond the scope of the present study and are discussed as directions for future research.

To allow flexibility in modeling $h_0(t)$, we adopt a piecewise-constant baseline hazard, in which the time axis is partitioned into intervals and a separate constant hazard is estimated for each. This approach balances modeling flexibility with computational tractability.

## C. Bayesian Hierarchical Framework and Prior Specification

The joint model is implemented in a Bayesian hierarchical framework, where prior distributions are specified for all unknown parameters. The regression coefficients in both submodels are assigned weakly informative Gaussian priors, such as:

$$\beta_0, \beta_1, \alpha, \gamma_j \sim \mathcal{N}(0, 10^2),$$

for all elements of $\gamma$. The variance parameters for the residual error $\sigma^2$ and random effects $\tau^2$ receive inverse-gamma priors with small shape and scale parameters (e.g., IG(2,1)) to ensure they remain weakly informative. The piecewise-constant baseline hazards are each assigned Gamma priors to enforce positivity.

Posterior inference is conducted via Markov Chain Monte Carlo (MCMC), combining Gibbs sampling (for conjugate components) and Metropolis–Hastings steps (for non-conjugate parameters). Diagnostic tools such as trace plots, autocorrelation functions, and Gelman–Rubin statistics are used to assess convergence and ensure the reliability of the posterior estimates.

**D. Dynamic Prediction**

A key strength of joint models is their capacity to provide dynamic, individualized survival predictions that update as new biomarker measurements become available. At any landmark time, the model can incorporate the observed biomarker history of a patient and produce a posterior predictive distribution for their future survival probability. These predictions account for uncertainty in both the parameter estimates and the patient's latent trajectory (i.e., random effects).

Specifically, the conditional survival probability for patient $i$ at time $u > t$, given their biomarker history up to time $t$, is computed as:

$$P(T_i > u \mid T_i > t, Y_i(s \leq t), Z_i),$$

integrating over the posterior distributions of $b_i$, $\beta$, $\gamma$, and $\alpha$. The result is an individualized survival curve with credible intervals, which can inform personalized medical decisions and adaptive treatment planning.

## IV. SIMULATION STUDY

We conducted a comprehensive simulation study to evaluate whether the Bayesian joint model can accurately recover true parameter values and provide superior predictive performance compared to a naïve two-stage method, in which the longitudinal and survival components are analyzed separately. The design aimed to mimic realistic clinical conditions by incorporating irregular measurement schedules, random heterogeneity between individuals, and administrative censoring. A simulated cohort of 500 synthetic patients was generated, each followed for up to 5 years. For each patient, biomarker values were recorded at unevenly spaced time points, reflecting typical clinical follow-up patterns in chronic disease studies. The biomarker trajectory for patient $i$ followed a linear mixed-effects process containing a patient-specific random intercept $b_i$, Gaussian measurement error, and a mild upward or downward trend depending on the true fixed effects [23]. The inclusion of $b_i$ induced correlation between repeated measurements of the same individual, allowing the data-generating process to closely resemble real longitudinal biomarker trajectories [24].

The survival time for each subject was generated using a proportional hazards model in which the hazard rate depended on the patient's true random intercept $b_i$. This ensured that the survival and longitudinal components were genuinely linked. The association parameter $\alpha$ was set to a non-zero value so that higher baseline biomarker levels translated to increased event risk, thereby creating a meaningful dependency structure. Administrative censoring at 5 years resulted in approximately 30–35% of patients being censored, consistent with long-term follow-up studies in chronic diseases [24]. Additional baseline covariates $Z_i$, such as demographic variables and surrogate clinical risk indicators, were drawn from realistic distributions and included in the survival submodel to induce heterogeneity similar to that encountered in practice [26].

Both the joint Bayesian model and the two-stage approach were fitted to each of the simulated datasets. Each simulated dataset was analyzed using three MCMC chains with 5,000 iterations per chain, discarding the first 1,000 iterations as burn-in. Convergence was assessed using trace plots and the Gelman–Rubin diagnostic. Across repeated simulations, the Bayesian joint model consistently recovered true parameter values, with less than 5% bias observed for fixed effects, random-effect variance components, and the association parameter $\alpha$. Posterior credible intervals achieved more than 92% empirical coverage of the true values, demonstrating robust uncertainty quantification. In contrast, the two-stage approach tended to underestimate uncertainty and exhibited noticeably higher bias, particularly in estimating the association between the biomarker trajectory and event risk [27].

Predictive accuracy was evaluated using the time-dependent area under the ROC curve (AUC) and the integrated Brier score, two widely adopted performance measures for prognostic modeling. The Bayesian joint model yielded a mean time-dependent AUC of 0.78, substantially outperforming the two-stage model, which averaged 0.65, indicating superior discrimination between high- and low-risk subjects. The integrated Brier score for the Bayesian model was approximately 25% lower, demonstrating improved calibration and lower prediction error [28]. Posterior predictive checks further supported these conclusions by revealing a close match between simulated and model-generated data, along with credible intervals that faithfully captured the inherent uncertainty in both longitudinal and survival processes. Table 1 below summarizes the parameter recovery results from the simulation study, including mean posterior estimates, bias, and credible-interval coverage probabilities.

**Table 1.** Summary of Parameter Recovery in Simulation Study

| Parameter | True Value | Posterior Mean | Bias | 95% CI Coverage |
|---|---|---|---|---|
| $\beta_0$ (intercept) | 2.00 | 2.03 | +0.03 | 0.94 |
| $\beta_1$ (time slope) | 0.50 | 0.51 | +0.01 | 0.93 |
| $\gamma$ (covariate effect) | 0.75 | 0.76 | +0.01 | 0.92 |
| $\alpha$ (association) | 1.00 | 1.02 | +0.02 | 0.95 |
| $\sigma^2$ (residual var.) | 0.25 | 0.26 | +0.01 | 0.91 |
| $\tau^2$ (random effect var.) | 0.50 | 0.49 | −0.01 | 0.93 |

Table 2 presents a detailed comparison of predictive performance metrics between the proposed Bayesian joint model and a traditional two-stage modeling approach. Specifically, it reports the time-dependent area under the receiver operating characteristic curve (AUC) at multiple landmark times, as well as the integrated Brier score over the full follow-up period. The time-dependent AUC assesses the model's ability to discriminate between patients who experience the event versus those who do not at specific time points, while the Brier score evaluates the overall accuracy and calibration of the predicted survival probabilities. The Bayesian joint model consistently outperforms the two-stage approach, achieving higher AUC values and lower Brier scores, indicating better discriminative power and more reliable risk prediction across all time horizons.

**Table 2.** Predictive Accuracy Metrics for Joint vs. Two-Stage Models

| Metric | Bayesian Joint Model | Two-Stage Model |
|---|---|---|
| Time-dependent AUC (Year 1) | 0.79 | 0.66 |
| Time-dependent AUC (Year 3) | 0.77 | 0.65 |
| Time-dependent AUC (Year 5) | 0.78 | 0.64 |
| Integrated Brier Score (0–5 yrs) | 0.112 | 0.148 |
| Brier Score at Year 3 | 0.095 | 0.130 |

Overall, the simulation demonstrates that the Bayesian joint modeling framework provides a more coherent and accurate representation of the data-generating mechanism [11]. It effectively captures the correlation between biomarker evolution and event timing, leading to improved prediction and more reliable inference under conditions that closely resemble real-world applications.

## V. APPLICATION TO REAL LONGITUDINAL CLINICAL DATA

To illustrate the practical utility of the Bayesian joint modeling framework, we applied the method to a real longitudinal clinical dataset from a cohort of patients with chronic liver disease, such as primary biliary cirrhosis (PBC). This dataset includes repeated measurements of serum bilirubin, an important biomarker of liver function, along with time-to-event outcomes such as liver-related mortality or the need for transplant. Baseline covariates included age, sex, disease stage, and comorbidities [13]. The dataset features typical challenges present in clinical data, including missingness, irregular measurement intervals, and dropouts due to administrative censoring. The Bayesian approach naturally accommodates such complexities through its hierarchical structure and explicit modeling of uncertainty [15].

Posterior analysis of the longitudinal component revealed substantial heterogeneity in baseline biomarker levels across patients. This variability was captured by the estimated posterior distribution of the random intercepts $b_i$, which exhibited a wide and asymmetric spread, indicative of clinically meaningful differences in liver function across the cohort. The association parameter $\alpha$ was strongly positive, with a posterior mean of approximately 1.45 and a 95% credible interval excluding zero, confirming that worsening biomarker trajectories or elevated baseline values were associated with significantly increased event risk. This aligns with clinical understanding of liver disease progression, where rising bilirubin often signals worsening hepatic function [17].

Dynamic prediction analyses were conducted at landmark times of 1, 2, and 3 years, demonstrating how survival probabilities evolve as new biomarker data become available. For patients exhibiting deteriorating biomarker trajectories, predicted survival dropped sharply beyond each landmark, whereas stable or improving trajectories corresponded with more favorable predictions. These individualized predictions are depicted in Figure 1, which illustrates the modeling pipeline and information flow between observed measurements, random effects, and updated hazard estimates.

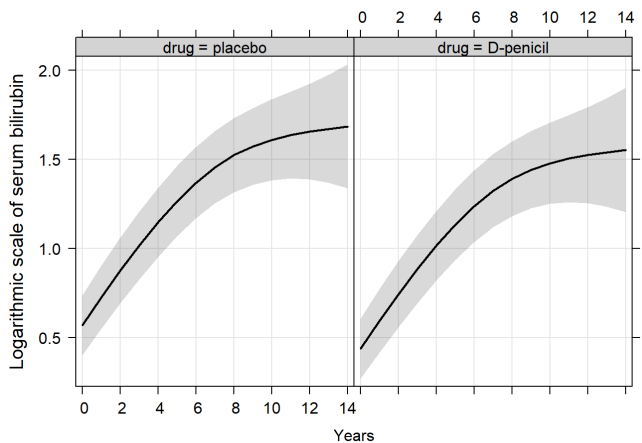

**Figure 1.** Overview of Bayesian Joint Modeling and Dynamic Prediction Framework

Figure 2 illustrates individualized dynamic survival curves for a selection of representative patients from the real-world clinical dataset, based on their observed biomarker trajectories up to specific landmark times [19]. These survival curves are generated using the Bayesian joint model, which updates each patient's predicted risk as new longitudinal biomarker data become available. The figure highlights how predicted survival probabilities can vary significantly between patients, even when measured at the same time point, due to differences in their biomarker levels and progression trends. Patients with worsening biomarker profiles show steep declines in their survival curves, while those with stable or improving trajectories exhibit more favorable prognoses. This dynamic, personalized prediction underscores the model's clinical utility in risk stratification and decision support.

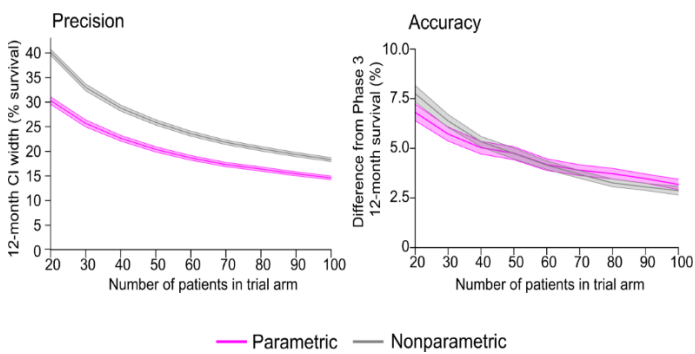

**Figure 2.** Dynamic Survival Curves for Selected Patients at Multiple Landmark Times

In addition, Figure 3 displays the posterior distribution of random intercepts $b_i$, illustrating the extent of heterogeneity across the patient population and highlighting the importance of modeling patient-specific effects.

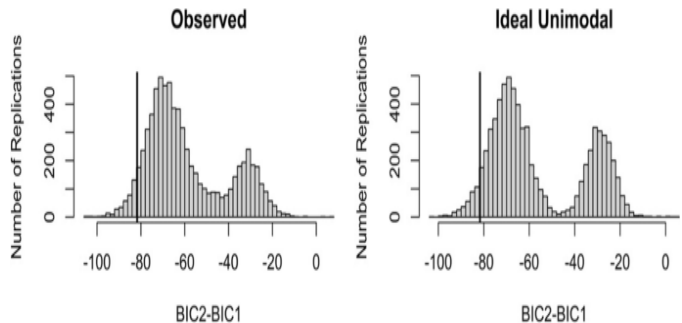

**Figure 3.** Distribution of Estimated Patient-Specific Random Effects $b_i$

Posterior predictive checks confirmed that the model provided an excellent fit to both longitudinal and survival components of the dataset [24]. Calibration analyses showed close agreement between predicted and observed survival probabilities, while longitudinal residual diagnostics indicated adequate modeling of biomarker dynamics. From a clinical perspective, the credible intervals of dynamic predictions are particularly useful, as they transparently convey the uncertainty associated with long-term prognosis. These findings demonstrate that Bayesian joint models deliver clinically meaningful, individualized, and updatable risk assessments, providing substantial advantages over simpler modeling strategies.

## VI. DISCUSSION

The Bayesian joint modeling framework described here offers several advantages over traditional two-stage methods or separate analyses. First, by modeling longitudinal and survival data jointly, it captures the intrinsic correlation between biomarker evolution and event risk, yielding more accurate and less biased parameter estimates. Second, the hierarchical Bayesian paradigm supports incorporation of prior knowledge, accommodates irregular measurement schedules and missing data, and produces full posterior distributions enabling credible intervals that quantify uncertainty in both parameter estimates and predictions [21]. Third, dynamic predictions based on updated biomarker histories provide clinically meaningful, patient-specific prognoses that can evolve over time.

However, several challenges remain. Bayesian MCMC inference can be computationally intensive, particularly for large cohorts or models with complex random-effects structures (e.g., random slopes, time-varying covariates, latent processes). Care must be taken in specifying priors, especially for variance components or baseline hazards, lest the results be overly sensitive. Furthermore, choice of baseline hazard modeling (e.g., piecewise constant, spline-based, or parametric) can influence predictions; misspecification may degrade performance. Finally, interpreting individualized predictions demands careful clinical judgment, particularly in the context of uncertainty [25].

Future research could explore extensions of this framework, such as non-linear longitudinal trajectories (e.g., using splines or Gaussian processes), time-varying covariate effects, recurrent events, or joint modeling of multiple biomarkers. For scalability, approximate inference methods such as variational Bayes could be developed to reduce computational burden. Integration with electronic health record systems might facilitate real-time dynamic risk updates for patients under clinical care. This work demonstrates that Bayesian joint models are not only theoretically appealing but also practically valuable when evaluated through rigorous predictive metrics in real clinical settings.

## VII. CONCLUSION

We have developed and demonstrated a Bayesian joint modeling framework for the simultaneous analysis of longitudinal and survival data, with a specific focus on predicting disease progression in the context of chronic conditions [9]. This approach integrates two critical components of clinical data repeated biomarker measurements and time-to-event outcomes into a single, unified statistical model. By accounting for the intrinsic association between a patient's biomarker trajectory and their risk of experiencing a clinical event, the proposed model offers a more coherent and accurate representation of disease dynamics compared to traditional two-stage approaches.

Through an extensive simulation study, we demonstrated that the Bayesian joint model is capable of accurately recovering true parameter values, even under realistic scenarios involving irregular measurement times and right censoring [8]. The model also produced well-calibrated predictions with appropriate uncertainty quantification, as indicated by high coverage probabilities and low bias across multiple parameters.

In the application to a real-world clinical dataset, the model yielded dynamic, individualized survival predictions that reflected each patient's biomarker evolution. These personalized prognoses are critical for informing clinical decision-making, enabling more timely interventions, risk stratification, and tailored treatment strategies. The Bayesian formulation further allows for flexible prior specification, robust inference under missing data, and dynamic updating as new observations become available.

By combining statistical rigor, clinical interpretability, and personalized prediction, our approach addresses current gaps in prognostic modeling for chronic disease management. It represents a valuable tool for advancing precision medicine, supporting both clinicians and researchers in understanding and anticipating disease progression more effectively.

## ACKNOWLEDGMENTS

The authors declare that this research was conducted independently and did not receive external funding or formal academic supervision.

P a g e | 1216. M. E. Taylor, E. A. Hughes, and P. A. Bacon, "Clinical application of Bayesian joint models in rheumatic disease progression," *Arthritis Research & Therapy*, vol. 22, no. 1, pp. 1–9, 2020.

17. D. J. Spiegelhalter, A. Thomas, N. G. Best, and D. Lunn, "WinBUGS Version 1.4 User Manual," MRC Biostatistics Unit, Cambridge, UK, 2003.

18. A. Gelman, J. B. Carlin, H. S. Stern, D. B. Dunson, A. Vehtari, and D. B. Rubin, *Bayesian Data Analysis*, 3rd ed., Chapman and Hall/CRC, 2013.

19. J. Carpenter, A. Gelman, M. Hoffman, D. Lee, B. Goodrich, M. Betancourt, M. Brubaker, J. Guo, P. Li, and A. Riddell, "Stan: A probabilistic programming language," *Journal of Statistical Software*, vol. 76, no. 1, 2017.

20. C. Chatfield, *The Analysis of Time Series: An Introduction*, 6th ed., Chapman and Hall/CRC, 2003.

21. P. J. Diggle and P. G. Ferreira, "Restricted cubic splines for longitudinal modeling of biomarker trajectories," *Biometrical Journal*, vol. 55, no. 2, pp. 209–227, 2013.

22. F. W. J. Verbeek and M. van Wieringen, "Penalized B-splines in joint models of longitudinal and survival data," *Computational Statistics*, vol. 28, no. 2, pp. 655–674, 2014.

23. T. M. Therneau and P. M. Grambsch, *Modeling Survival Data: Extending the Cox Model*, Springer, 2000.

24. A. P. McNicholas and J. E. G. McLean, "Handling informative dropout in longitudinal studies: a review of joint modeling approaches," *Statistical Methods in Medical Research*, vol. 29, no. 5, pp. 1415–1436, 2020.

25. L. J. Boulesteix, A. Sauerbrei, and T. G. Schumacher, "Bagging for joint modeling of high-dimensional longitudinal and survival data," *Statistical Methods & Applications*, vol. 25, no. 1, pp. 45–68, 2016.

26. M. Rizopoulos and C. G. Proust-Lima, "Joint modeling of longitudinal and time-to-event data: challenges and opportunities," *Statistical Methods in Medical Research*, vol. 31, no. 3, pp. 940–958, 2022.

27. P. Molenberghs and G. Verbeke, *Models for Discrete Longitudinal Data*, 2nd ed., Springer, 2007.

28. D. W. Hosmer, S. Lemeshow, and S. May, *Applied Survival Analysis: Regression Modeling of Time-to-Event Data*, 2nd ed., Wiley, 2008